\documentclass[11pt]{article}
\usepackage{amsmath}
\usepackage{amssymb}
\usepackage{graphicx}
\usepackage{cite}

\title{\bf\Large \boldmath$\Phi$-Values in Protein Folding Kinetics Have Energetic and Structural Components}
\author{\hspace*{-1cm}Claudia Merlo$^1$, Ken A.\ Dill$^2$, and Thomas R.\ Weikl$^1$\\[0.2cm]
\hspace*{-1cm}\small $^1$Max Planck Institute of Colloids and Interfaces, Theory Division,
14424 Potsdam, Germany\\
\hspace*{-1cm}\small $^2$Department of Pharmaceutical Chemistry, 
University of California, San Francisco,\\[-0.1cm] 
\hspace*{-1cm}\small California 94143-2240, USA} 
\date{}
\begin{document}
\maketitle

\begin{abstract}
$\Phi$-values are experimental measures of how the kinetics of protein folding is changed by single-site mutations.  $\Phi$-values measure {\it energetic} quantities, but are often interpreted in terms of the {\it structures} of the transition state ensemble.  Here we describe a simple analytical model of the folding kinetics in terms of the formation of protein substructures. The model shows that $\Phi$-values have both structural and energetic components. In addition, it provides a natural and general interpretation of ``nonclassical" $\Phi$-values (i.e., less than zero, or greater than one).  The model reproduces the $\Phi$-values for 20 single-residue mutations in the $\alpha$-helix of the protein CI2, including several nonclassical $\Phi$-values, in good agreement with experiments.
\end{abstract}

\subsection*{Introduction}

The folding kinetics of small single-domain proteins has been widely studied by single-site mutagenesis \cite{otzen94,itzhaki95,villegas98,chiti99,martinez99,riddle99,ternstrom99,kragelund99,kim00,mccallister00,hamill00,fowler01,jaeger01,otzen02,northey02,garcia04}.  The central quantity in these studies, the $\Phi$-value, is given by \cite{matouschek89,fersht99}
\begin{equation}
\Phi= \frac{R T\ln( k_{\text{wt}}/k_{\text{mut}})}{\Delta G_{N}} \label{phi}
\end{equation}
where $k_{\text{wt}}$ and $k_{\text{mut}}$ are the folding rates of the wildtype and mutant protein, and $\Delta G_{N}$ is the change of the protein stability upon mutation. The stability $G_{N}$ of a protein is the free energy difference between the native state $N$ and the denatured state $D$.

There are several theoretical studies of $\Phi$-values and transition states. The thermal unfolding kinetics of CI2 has been extensively studied in MD simulations \cite{li94,li96,daggett96,ladurner98,lazaridis97,kazmirski01}. Here, the transition state is defined as a ``small ensemble of structures populated immediately prior to the onset of a large structural change'' \cite{li96} in the unfolding trajectories. Other groups have considered statistical mechanical or Go-type models 
\cite{alm99,munoz99,galzitskaya99,portman01,alm02,kameda03,clementi00,hoang00,li01,karanicolas02,kaya03,weikl04}.  In some of these models, transition states are identified as free energy maxima along a folding reaction coordinate,  or as  free energy saddle points if two or more degrees of freedom are used for the reaction coordinate. More recent approaches define the transition state ensemble from experimental $\Phi$-values by using these $\Phi$-values as restraints in simulations \cite{vendruscolo01,lindorff03,lindorff04}. Each of these definitions of transition state, while plausible, is nevertheless based on one or more ad hoc premises.

Using classical transition state theory, the folding rate is proportional to $\exp[-G_T /RT]$ where $G_{T} = G_{\text{transition state}} - G_{\text{denatured state}}$ is the free energy difference between the transition state ensemble and the denatured state. Possible changes in the prefactor of this proportionality relation upon mutation are usually neglected. Thus, $\Phi=\Delta G_{T}/ \Delta G_{N}$. In this way, $\Phi$-values measure the {\em energetic} consequences of mutations on the transition state ensemble relative to the native state.

A central question is whether $\Phi$-values also give {\em structural} information about the transition state ensemble \cite{fersht99,hammond55,matthews93}. In the traditional interpretation, $\Phi=1$ is taken to indicate that the mutated residue has native-like structure in its transition state ensemble (TSE), while $\Phi=0$ is taken to indicate that the mutated residue is not structured in the TSE.  Typically, experiments give $\Phi$-values that are fractional, with values between 0 and 1, apparently indicating partial native-like structural character of the residue in the TSE.

However, there are three problems with this traditional structural interpretation. First, $\Phi$-values are sometimes ``nonclassical"; they can be less than zero or larger than one.  In the traditional view, such values are impossible, implying a transition state that is more denatured than D or more native than N; hence there is some controversy about how such $\Phi$-values should be interpreted. Second, a given sequence position can have very different $\Phi$-values, depending on which amino acid is substituted there, leading to the question of whether such energetic changes always have a simple structural interpretation.

\begin{figure}
\vspace{-2cm}
\begin{center}
\resizebox{\linewidth}{!}{\includegraphics{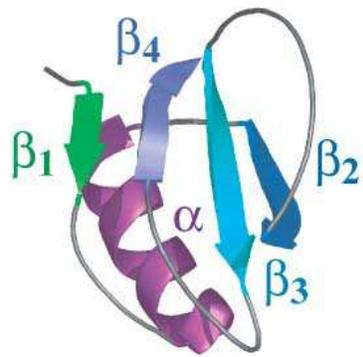}}
\end{center}
\vspace{0cm}

\caption{The native structure of CI2 consists of a four-stranded $\beta$-sheet packed against an $\alpha$-helix (PDB file 1COA). \label{ci2structure}} 
\end{figure}

Third, there is a problem of continuity: two residues that are neighbors in the chain are sometimes observed to have very different $\Phi$-values.  A structural interpretation of this would be that there can be sharp boundaries between native-like and non-native-like structure in the TSE, which seems implausible.  For example, the protein CI2 consists of an $\alpha$-helix packed against a four-stranded $\beta$-sheet (see Fig.~\ref{ci2structure}). Twenty single residue-mutations have been studied in the $\alpha$-helix of CI2, giving $\Phi$-values ranging over the full spectrum from -0.35 to 1.25.  Even though helix formation is usually regarded as fast and cooperative, these results would seem to imply that this helix does not form as a single cooperative unit: parts are folded and parts are not in the TSE.  It is not clear whether these are problems of experimental errors, or problems in the traditional model that is used to interpret $\Phi$-values.

Is there a more physical way to interpret the formation of protein substructures that comprise the TSE of protein folding?  We develop here a model.  We first consider the simplest subdivision of the protein: into one $\alpha$-helical substructure and one $\beta$-sheet substructure.  Because of its simplicity, the model can be solved analytically and exactly.  We then generalize this model to apply to CI2.  Despite its simplicity, this model reproduces the experimental $\Phi$-values in CI2 with a correlation coefficient of 0.85, including some of the nonclassical $\Phi$-values.  A key conclusion is that it is not sufficient to interpret $\Phi$-values solely in terms of structures.  A $\Phi$-value can, however, be decomposed into structural and energetic components.

\subsection*{The Dynamics}

Our approach has two aspects: (1) the model, which expresses the relative free energies of the various substructures of the protein as it folds, and (2) the dynamics of the model.  We first describe our treatment of the dynamics.  To simplify the notation, we define here the free energy $G_n$ of each partially folded state $n = 1, 2, 3, \ldots$, and the dimensionless free energy $g_N\equiv G_N/RT$, with respect to the fully denatured state in which none of the substructures is formed. Thus the denatured state is the reference, defined as having zero free energy. The transition rate from any state $m$ to state $n$ is given by
\begin{equation}
w_{nm}=\frac{1}{t_o} \left(1+e^{g_{n}-g_{m}}\right)^{-1} \label{transrates} 
\end{equation}
provided the states $n$ and $m$ are connected via a single step in which only one substructure folds or unfolds \cite{weikl04}.  For other transitions, the transition rates are zero. Here, $t_o$ is a reference time scale.\footnote{The transition rates obey detailed balance $w_{nm} P_{m}^e = w_{mn} P_{n}^e$ where $P_{n}^e\sim \exp[-G_n/(R T)]$ is the equilibrium weight for the state $n$. Detailed balance ensures that the system ultimately reaches thermal equilibrium.}

The folding kinetics is described by the master equation 
\begin{equation}
\frac{\text{d}{\boldsymbol P}(t)}{\text{d} t} = -{\boldsymbol W} \boldsymbol{P}(t) 
\end{equation}
The elements of the vector $\boldsymbol{P}(t)$  are the probabilities $P_n(t)$ that the protein is in state $n$ at time $t$, and the matrix elements of $\boldsymbol{W}$ are given by $W_{nm} = -w_{nm}$ for $n\neq m$ and $W_{nn} = \sum_{m\neq n} w_{mn} $. The general solution of the master equation is
\begin{equation}
\boldsymbol{P}(t)=\sum_{\lambda} c_{\lambda} \boldsymbol{Y}_{\lambda} \exp[-\lambda t] \label{gensol} 
\end{equation}
which is expressed in terms of the eigenvalues $\lambda$ and eigenvectors $\boldsymbol{Y}_\lambda$ of the matrix $\boldsymbol{W}$. The prefactors $c_{\lambda}$ depend on the initial conditions at time $t=0$. 

The eigenvalues represent relaxation rates. It can be shown that one eigenvalue is zero, corresponding to the equilibrium distribution, while all other eigenvalues are positive \cite{vanKampen}. For $t\to\infty$, the probability vector $\boldsymbol{P}(t)$ tends to $c_o \boldsymbol{Y}_o$ where  $\boldsymbol{Y}_o$ is the eigenvector with eigenvalue 0.

\subsection*{The Model: Two Substructures}

The dynamics above is applicable to any model of the protein, its substructures, and their relative free energies.  Here we first apply the dynamics to the simplest possible model of the substructures of CI2.  There are four states in the model: (1) the denatured state $D$, in which neither the helix nor the sheet is formed; (2) a partially folded state $\alpha$, in which only the helix is formed; (3) a partially folded state $\beta$, in which only the $\beta$-sheet is formed; and (4) the native state $N$, in which both the helix and sheet are formed and packed against each other.

In this simple four-state model, the energy landscape is characterized by the dimensionless free energy differences $g_\alpha$, $g_\beta$, and $g_N$ of the states $\alpha$, $\beta$, and $N$, each taken with respect to the denatured state $D$, which is defined as having zero free energy. 

The folding kinetics of this model can be solved exactly by determining the eigenvalues $\lambda$ and eigenvectors $\boldsymbol{Y}_\lambda$ of the matrix  $\boldsymbol{W}$. Since this model has four states, $\boldsymbol{W}$ is a $4\times 4$ matrix. In units of $1/t_o$, the eigenvalues are given by $\lambda = 0$, $1-q$, $1+q$, and 2 where
\begin{equation}
q=\frac{1-e^{g_{N}-g_\alpha-g_\beta}}{\sqrt{(1+e^{-g_\alpha})(1+e^{-g_\beta})
   (1+e^{g_{N}-g_\alpha})(1+e^{g_{N}-g_\beta})}} \label{q}
\end{equation}
Since we have $-1< q< 1$, the three nonzero eigenvalues are positive and describe the relaxation to the equilibrium state of the model (see eq.~(\ref{gensol})). The equilibrium state simply is  $c_o \boldsymbol{Y}_o$ where  $\boldsymbol{Y}_o$ is the eigenvector with eigenvalue 0.

This model exhibits two-state folding kinetics under two conditions.  First, the native state must be stable: the free energy $g_N$ of the native state must be significantly smaller than the free energies of the other three states.  Under such folding conditions, the equilibrium native state will be more populated than the other three states. Second, the intermediate states $\alpha$ and $\beta$ must have positive free energies, relative to D, so that the system will have a kinetic barrier, which is required to achieve single-exponential dynamics.  

Under these two conditions, the three Boltzmann weights $e^{g_{N}-g_\alpha-g_\beta}$, $e^{g_{N}-g_\alpha}$, and $e^{g_{N}-g_\beta}$ in eq.~(\ref{q}) are much smaller than 1, and also much smaller than $e^{-g_\alpha}$ and $e^{-g_\beta}$.  Therefore, these three Boltzmann weights can be neglected. We set them to zero. The factor $q$ in eq.~(\ref{q}) then simplifies to 
\begin{equation}
q\simeq \frac{1}{\sqrt{(1+e^{-g_\alpha})(1+e^{-g_\beta})}}
\end{equation}
For large barrier energies $g_\alpha$ and $g_\beta$, we have $e^{-g_\alpha}\ll 1$ and $e^{-g_\beta}\ll 1$, and therefore $(1+e^{-g_\alpha})(1+e^{-g_\beta})\simeq (1+e^{-g_\alpha} +e^{-g_\beta})$. If we now use the expansion $(1+x)^{-1/2}\simeq 1- x/2$ with $x =e^{-g_\alpha} +e^{-g_\beta} \ll 1$, the smallest nonzero relaxation rate, or folding rate, $k\equiv 1-q$ is given by,
\begin{equation}
k \simeq \frac{1}{2} \left(e^{-g_\alpha} +e^{-g_\beta}\right) \label{foldingRate}
\end{equation}
The folding rate $k$ is much smaller than the other two relaxation rates $1+q$ and 2.  In that case, these two fast relaxations constitute an initial `burst phase' and the model otherwise gives two-state single-exponential folding behavior with slowest rate $k$ (see eq.~(\ref{gensol})).  The folding rate $k$ simply is the sum of the rates for the two possible folding routes: one in which $\alpha$ forms first and the other in which $\beta$ forms first. The factor 1/2 in the equation above arises because a molecule, after reaching one of the barrier states $\alpha$ or $\beta$, either falls back to $D$ or falls forward to $N$, with almost equal probability. 

Using this model, we now explore the effects of mutations.  Consider a mutation within the $\alpha$-helix. The free energy of the helix will change from $g_\alpha \to g_\alpha + \Delta g_\alpha$ and the free energy of the native state will change from $g_N \to g_N + \Delta g_N$.  In contrast, $g _\beta$ is not affected by the mutation. The folding rate of the mutant will be $k_\text{mut}=k(g_\alpha+\Delta g_\alpha,g_\beta)$ with $k$ given by eq.~(\ref{foldingRate}).  For small perturbations $\Delta g_\alpha$, we have $\ln k_\text{wt} - \ln k_\text{mut}\simeq -(\partial \ln k/\partial\, g_\alpha)\Delta g_\alpha$. For mutations in the $\alpha$-helix, the $\Phi$-value defined in eq.~(\ref{phi}) thus has the general form
\begin{equation}
\Phi = \chi_\alpha\;\frac{\Delta g_\alpha}{\Delta g_N}
\label{phi_alpha}
\end{equation}
with
\begin{equation}
\chi_\alpha = -\frac{\partial \ln k}{\partial \, g_\alpha}  =\frac{e^{-g_\alpha}}{e^{-g_\alpha}+e^{-g_\beta}}  
\end{equation}
Hence, the $\Phi$-value is a product of two terms: a {\em structural} factor $\chi_\alpha$, and an {\em energetic} factor $\Delta g_{\alpha}/{\Delta g_N}$. The term $\chi_\alpha$ describes the {\em fractional structure formation} of the $\alpha$-helix within the TS ensemble. In this example, the TSE consists of the two barrier states $\alpha$ and $\beta$ on the two parallel folding routes. $\chi_\alpha$ ranges between 0 and 1. We have $\chi_\alpha = 1$ for $g_\alpha \ll g_\beta$ when the state $\alpha$ dominates the TSE, and $\chi_\alpha = 0$ when $\beta$ dominates the TSE.

Whereas $\chi_{\alpha}$ gives structural information, the second term, $\Delta g_{\alpha}/{\Delta g_N}$, can take on either negative or positive values. This term thus accounts for nonclassical $\Phi$-values smaller than 0 or larger than 1.  In the simplest case, we have $\Delta g_N = \Delta g_\alpha + \Delta g_{\alpha\beta}$. Here, $\Delta g_{\alpha\beta}$ is the free energy change for a tertiary contact between the $\alpha$-helix and the $\beta$-sheet, for example.  In that case, negative $\Phi$-values arise when $\Delta g_{\alpha\beta}$ is larger in magnitude and opposite in sign to that of  $\Delta g_\alpha$.  That is, a negative $\Phi$-value is predicted when a helical mutation also has a counteracting and larger effect on a tertiary contact.  Correspondingly, $\Phi > 1$ occurs when two conditions are met: (1)  $\Delta g_{\alpha\beta}$ is opposite in sign, but smaller in magnitude than $\Delta g_\alpha$,   
and (2) $\chi_\alpha$ is sufficiently large.  This explanation of nonclassical $\Phi$-values may also rationalize why more $\Phi$-values are negative than larger than 1 [42]. If $g_\alpha$ and $g_{\alpha\beta}$ have a similar magnitude, it should be more difficult to satisfy the latter two conditions than the former one.

However, our model is rather general and captures also that nonclassical $\Phi$-values can arise from shifts in the free energy of the denatured state. For example, if a mutation only lowers the free energy of the denatured state, we have $\Delta g_{\alpha}>0$ and $\Delta g_N<0$, which gives a negative $\Phi$-value according to eq.~(\ref{phi_alpha}).  In contrast, the traditional structural interpretation of $\Phi$-values fails if mutations shift the free energy of the denatured state \cite{fersht04}.

In this simple example, a mutation in the $\alpha$-helix affects only a single structural element formed in the TSE: the $\alpha$-helix itself. In general, mutations may affect several microstructures of the TSE.  A generalization of eq.~(\ref{phi_alpha}) then is $\Phi = (\sum_i \chi_i \Delta g_i)/\Delta g_N$ with $\chi_i = -(\partial \ln k)/(\partial g_i)$, provided the free energies $g_i$ of the microstructures are additive.

\subsection*{Mutations in the $\alpha$-helix of CI2}

\begin{figure}[t]
\hspace*{-0.4cm}
\vspace{1cm}

\begin{center}
\hspace*{-0.4cm}
\resizebox{0.5\linewidth}{!}{\includegraphics{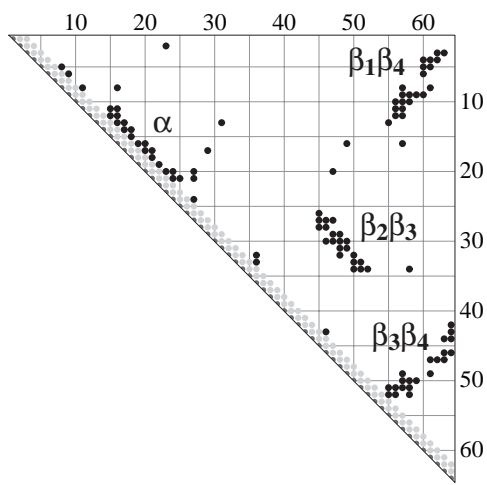}}
\end{center}

\caption{Contact matrix of CI2. Each black dot represents a contact between two amino acids in the native structure, with a distance of less than 6 \AA\ between the C$_\alpha$ or  C$_\beta$ atoms of the amino acids. The four large clusters of contacts correspond to the main structural elements of CI2: the $\alpha$-helix and the $\beta$-strand pairings $\beta_2\beta_3$, $\beta_3\beta_4$, and $\beta_1\beta_4$. The few `isolated' contacts either represent turns or tertiary interactions of $\alpha$-helix and $\beta$-sheet. \label{ci2contactMap}}
\end{figure}

To model the folding kinetics of CI2, we must consider at least four substructural units: the $\alpha$-helix, and the three strand pairings $\beta_2\beta_3$, $\beta_3\beta_4$, and  $\beta_1\beta_4$. These substructures correspond to contact clusters on the native contact map of CI2 (see Fig.~\ref{ci2contactMap}). The model energy landscape of CI2 therefore is more complex than the landscape of the simple four-state model given above. However, under two assumptions, eq.~(\ref{phi_alpha}) also holds for the helix of CI2. These assumptions are: (1) the helix is either fully formed or not formed in each of the states of the transition state ensemble, and (2) the helix does not form tertiary contacts in the transition state ensemble. Under these assumptions, the free energy contribution of the helix to a state of the transition state ensemble (in which the helix is formed) simply is $g_{\alpha}$, and then $\chi_\alpha$ has the same interpretation as above.\footnote{These two assumptions are clearly simplifying. Based on unfolding simulations, Daggett et al.\ \cite{daggett96} argue for a crucial tertiary interaction between the residues Ala16 of the $\alpha$-helix and Ile49 of the $\beta$-sheet in the transition state ensemble of CI2.  In contrast, Lazaridis and Karplus \cite{lazaridis97} found that ``the  the number of contacts made by the Ala side chain [in the TSE] \ldots depend[s] primarily on the presence of the helix and not on interactions with $\beta$-strands.''}

\begin{figure}

\begin{center}
\hspace*{-2cm}
\resizebox{0.8\linewidth}{!}{\includegraphics{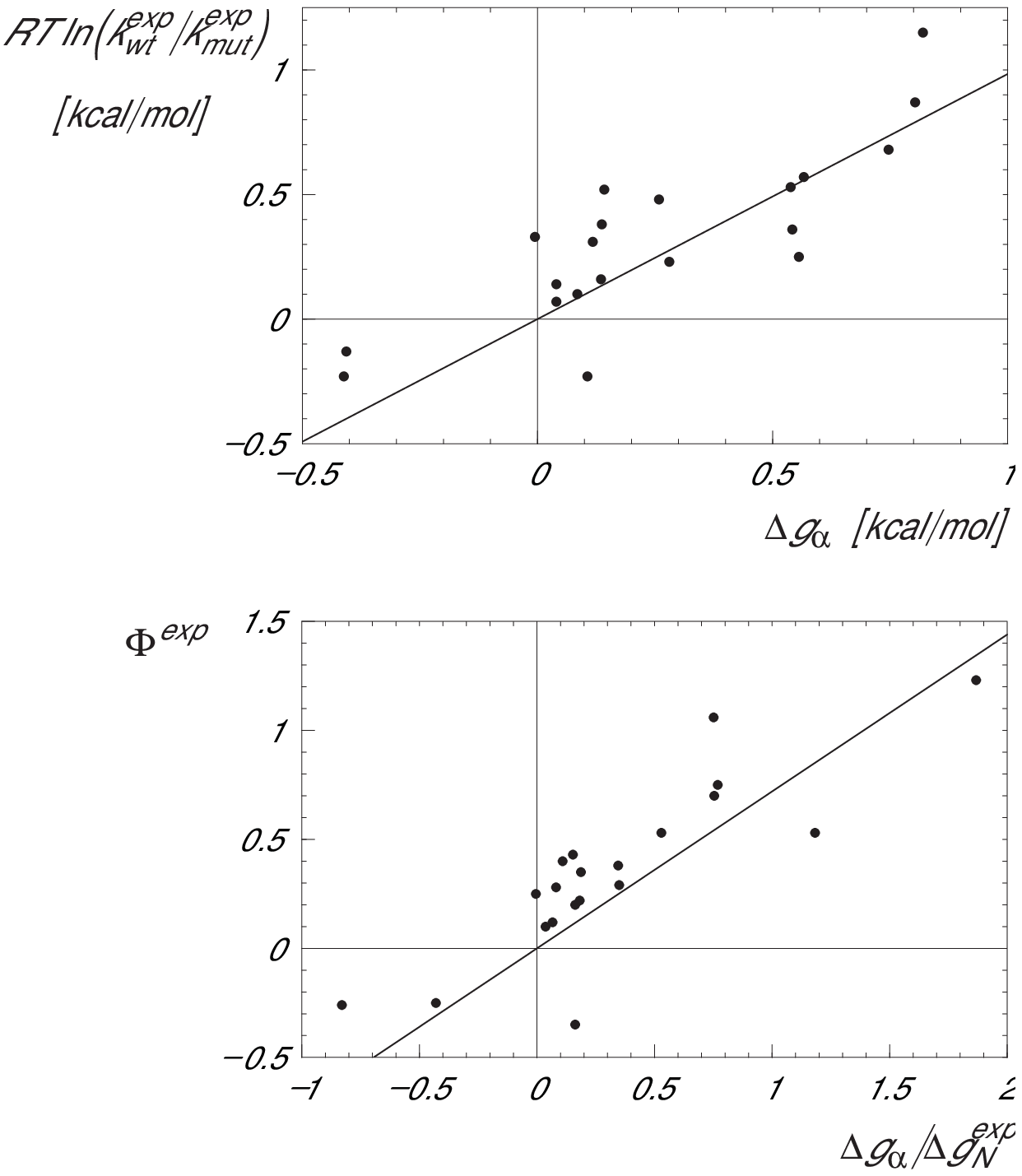}}
\end{center} 
 
\caption{Correlation analysis for mutations in the CI2 helix. (Top) $\ln(k^{\text{exp}}_{\text{wt}}/k^{\text{exp}}_{\text{mut}})$  versus $\Delta g_{\alpha}= \ln (P_{\alpha}^{\text{wt}}/P_{\alpha}^{\text{mut}})$ estimated from helicities $P_\alpha$ predicted by AGADIR (see Table 1). The correlation coefficient $r$ is 0.83, and the slope of the fitted line through the origin is 0.98. The slope of this line is an estimate for the parameter $\chi_\alpha$ of eq.~(\ref{phi_alpha}). For subsets of the data generated by deleting up to two data points, the correlation coefficient $r$ varies from 0.77 to 0.93, and the linear slope varies from 0.87 to 1.07. (Bottom) $\Phi_{\text{exp}}$ versus $\Delta g_{\alpha}/\Delta g_{N}^\text{exp}$. The correlation coefficient $r$ is 0.85, and the slope of the fitted line through the origin is 0.71. For data subsets generated by deleting up to two data points, $r$ varies from 0.79 to 0.90, and the slope varies from 0.64 to 0.90.}
\label{ci2helix}
\end{figure}

To test eq.~(\ref{phi_alpha}), we consider the 20 single-residue mutations in the CI2 helix \cite{itzhaki95}. We estimate the change in intrinsic helix stability $\Delta g_{\alpha}$ from helicities predicted by the program AGADIR \cite{munoz94a,munoz94b,lacroix98} (see Table 1). The experimentally measured change in folding rate for these mutations, $\log(k_\text{wt}^{\text{exp}}/ k_\text{mut}^{\text{exp}})$,  correlates with $\Delta g_{\alpha}$ with a coefficient $r=0.83$, and the experimentally determined $\Phi$-values correlate with $\Delta g_{\alpha}/\Delta g_{N}^\text{exp}$ with $r=0.85$ (see Fig.~\ref{ci2helix}). According to eq.~(\ref{phi_alpha}), the change in $\log k$ is proportional to $\Delta g_{\alpha}$, and the $\Phi$-values are proportional to $\Delta g_{\alpha}/\Delta g_{N}^\text{exp}$, both with proportionality constant $\chi_\alpha$. From the two linear fits shown in Fig.~\ref{ci2helix}, we obtain the estimate $\chi_\alpha = 0.88\pm 0.12$. We have estimated the errors for $\chi_\alpha$ using a jackknife method in which up to two data points are deleted randomly from the data set (see figure caption). This estimate for $\chi_\alpha$ indicates that the helix is almost fully formed in the transition state ensemble. In agreement with this interpretation, MD unfolding simulations indicate that a fraction of $0.91\pm0.14$ of the helical residues are structured in the transition state ensemble \cite{daggett96}.

\subsection*{Discussion}

Our model gives a physical explanation for nonclassical $\Phi$-values, but an alternative explanation is in terms of experimental errors.  S\'anchez and Kiefhaber \cite{sanchez03} have observed that mutations with nonclassical $\Phi$-values often have relatively small changes $\Delta g_{N}$ in stability.  Since $\Delta g_{N}$ appears in the denominator of the expression for $\Phi$, it means that nonclassical $\Phi$-values can arise when a mutation has little effect on the protein stability.  S\'anchez and Kiefhaber argue that unavoidable experimental errors may be responsible for the unusual $\Phi$-values, and that $\Phi$-values for mutations with $\Delta g_{N}<1.7$ kcal/mol are unreliable. Others have argued that this error threshold should be considerably smaller, around  0.6 kcal/mol \cite{fersht04,garcia04}. The analysis of S\'anchez and Kiefhaber is based on the assumption that different mutations at a given residue position should lead to the same `true' $\Phi$-value for this residue position.  Our model gives a different interpretation.  In our model, different mutations at a given position can affect the energy landscape in different ways.  For example, we believe E14Q in the CI2 helix may affect the helicity significantly, while E14D does not (see Table 1).

Our model can explain {\em isolated} nonclassical $\Phi$-values, such as the four in the $\alpha$-helix of CI2 (see Table 1).  They are ``isolated" insofar as they are interspersed among classical $\Phi$-values within a local region of the protein.  There are other cases in which nonclassical $\Phi$-values are {\em clustered together} within a given region of the protein. In the second $\alpha$-helix of ACBP for example, 7 $\Phi$-values are clearly negative, while the other 6 $\Phi$-values are close to 0.  Previously, clustered nonclassical $\Phi$-values have been explained in terms of parallel flow processes on slightly more complex energy landscapes than we considered here \cite{ozkan01}.  That is, mutations that destabilize a particular substructure can cause a backflow on the energy landscape into faster flow channels, leading to an increase in the folding rate and negative $\Phi$-values.

We have  considered here the $\alpha$-helix of CI2  to illustrate our structural interpretation of $\Phi$-values. One reason is that the helix is very well characterized, i.e.~a large number of $\Phi$-values is available. Another reason is that these $\Phi$-values cover a wide range of possible values, from \mbox{-0.35} to 1.23. Two other well-characterized helices are the $\alpha$-helices of protein L \cite{kim00} and protein G \cite{mccallister00}. 15 single-residue mutations have been considered in the protein L helix. One of the $\Phi$-values is -0.39, whereas the others span a rather narrow range from -0.05 to 0.28 \cite{kim00}. Similarly, one out of 9 $\Phi$-values for the helix of protein G is -0.81, whereas the others range from 0.05 to 0.55. In both cases, our model reproduces the clearly negative, nonclassical $\Phi$-value, which leads to relatively high correlation coefficients of 0.58 and 0.81 between the experimental and theoretical $\Phi$-value distributions. But since the other $\Phi$-values lie in  a rather narrow range, the statistical uncertainties from experimental and modeling errors are high and $\chi_\alpha$ can not be determined reliably.

\subsection*{Summary}

$\Phi$-values give information about the routes of protein folding.  The central question is: What information do they give?  Previous modeling has been limited in certain ways.  First, some models treat only topological aspects of folding, and therefore cannot explain how single-site mutations can have the large effects on folding rates that are often observed.  Second, current models usually make some plausible, but ad hoc, assumption about folding routes, transition states, and reaction coordinates.  Protein folding is sufficiently different than simpler reactions that some of these assumptions are not likely to be valid.  In particular, $\Phi$-values are often assumed to reflect only structural information about transition states.  Here we present a more rigorous approach for interpreting $\Phi$-values, and we show that $\Phi$-values have both structural and energetic components.  We show that our approach gives a consistent interpretation of mutational experiments on the CI2 helix.

\newpage

\begin{table}

\hspace*{0.8cm}
Table 1: Data for single-residue mutations in the $\alpha$-helix of CI2\\ 
\begin{center}
\begin{tabular}{cccccc}
 mutation & $RT \ln(k^{\text{exp}}_{\text{wt}}/k^{\text{exp}}_{\text{mut}})$ & $\Delta g_{N}^{\text{exp}}$  & $\Phi^{\text{exp}}$ & $\Delta g_{\alpha}$ & $ \Delta g_{\alpha}/\Delta g_{N}^{\text{exp}}$ \\ 
 \hline
S12G & 0.23 & 0.8 &  0.29 & 0.28 &  0.35\\ 
 S12A & 0.38 & 0.89 & 0.43 & 0.14 & 0.16 \\ 
 E14Q & 0.36 & 0.29 & 1.23 & 0.54  & 1.86 \\ 
 E14D & 0.10 & 0.52 & 0.2 & 0.08 & 0.15\\ 
 E14N & 0.53 & 0.7 & 0.75 & 0.54 & 0.77 \\ 
 E15Q & 0.25 & 0.47 & 0.53 & 0.56 & 1.19 \\ 
 E15D & 0.16 & 0.74 & 0.22 & 0.13 & 0.18 \\ 
 E15N & 0.57 & 1.07 & 0.53 & 0.57 & 0.53 \\ 
 A16G & 1.15 & 1.09 & 1.06 & 0.82 & 0.75 \\ 
 K17A & 0.14 & 0.49 & 0.28 & 0.04 & 0.08 \\ 
 K17G & 0.87 & 2.32 & 0.38 & 0.80 & 0.34 \\ 
 K18G & 0.68 & 0.99 & 0.7 & 0.75 & 0.76 \\ 
 V19A & -0.13 & 0.49 & -0.26 & -0.41 & -0.84 \\ 
 I20V & 0.52 & 1.3 & 0.4 & 0.14 & 0.11 \\ 
 L21A & 0.33 & 1.33 & 0.25 & -0.01 & -0.01 \\ 
 L21G & 0.48 & 1.38 & 0.35 & 0.26 & 0.19 \\ 
 Q22G & 0.07 & 0.6 & 0.12 & 0.04 & 0.07 \\ 
 D23A & -0.23 & 0.96 & -0.25 & -0.41 & -0.43 \\ 
 K24A & -0.23 & 0.65 & -0.35 & 0.11 & 0.17 \\ 
 K24G & 0.31 & 3.19 & 0.1 & 0.12 & 0.04
\end{tabular}
\end{center}
Experimental data for folding rates $k^{\text{exp}}_{\text{wt}}$ and $k^{\text{exp}}_{\text{mut}}$ of wildtype and mutants, stability changes $\Delta g_{N}^{\text{exp}}$, and $\Phi$-values are from Itzhaki et al.\cite{itzhaki95}. The change 
$\Delta  g_{\alpha}= \ln (P_{\alpha}^{\text{wt}}/P_{\alpha}^{\text{mut}})$ in the `intrinsic helix stability' $g_{\alpha}$ is estimated from helicities $P_{\alpha}$ predicted by AGADIR\cite{munoz94a,munoz94b,lacroix98}.  The wildtype sequence of the 13-residue helix is SVEEAKKVILQDK. Helicities have been calculated at the experimental temperature 298 K, pH 6.25, and ionic strength 0.03 mol, with acetylated N-terminus and amidated C-terminus of the peptide to avoid terminal charges. The energetic quantities  $RT \ln(k^{\text{exp}}_{\text{wt}}/k^{\text{exp}}_{\text{mut}})$,  $\Delta g_{N}^{\text{exp}}$,   and $\Delta g_{\alpha}$ are given in units of kcal/mol.
\end{table}

\clearpage

\clearpage

\clearpage

\begin{thebibliography}{99}

\small

\bibitem{otzen94} Otzen, D.E., Itzhaki, L.S., elMasry, N.F., Jackson, S.E., \& Fersht, A.R. (1994) {\it Proc.\ Natl.\ Acad.\ Sci.\ USA} {\bf 91}, 10422-10425.

\bibitem{itzhaki95} Itzhaki, L.S., Otzen, D.E., \& Fersht, A.R. (1995) {\it J.\ Mol.\ Biol.} {\bf 254}, 260-288. 

\bibitem{villegas98} Villegas, V., Martinez, J.C., Aviles, F.X.,  \& Serrano, L. (1998) {\it J.\ Mol.\ Biol.} {\bf 283}, 1027-1036.

\bibitem{chiti99} Chiti, F., Taddei, N., White, P.M., Bucciantini, M., Magherini, F., Stefani, M., \& Dobson, C.M.  (1999) {\it Nat.\ Struct.\ Biol.} {\bf 6}, 1005-1009. 

\bibitem{martinez99} Martinez, J.C.,  \& Serrano, L. (1999) {\it Nature Struct.\ Biol.} {\bf 6}, 1010-1016. 

\bibitem{riddle99} Riddle, D.S., Grantcharova, V.P., Santiago, J.V., Alm, E., Ruczinski, I.,  \& Baker, D. (1999)  {\it Nature Struct.\ Biol.} {\bf 6}, 1016-1024. 

\bibitem{ternstrom99} Ternstr\"om, T., Mayor, U., Akke, M., \& Oliveberg, M. (1999)  {\it Proc.\ Natl.\ Acad.\ Sci.\ USA} {\bf 96}, 14854-14859. 

\bibitem{kragelund99} Kragelund, B.B., Osmark, P., Neergaard, T.B., Schiodt, J., Kristiansen, K., Knudsen, J.,  \& Poulsen, F.M. (1999) {\it Nature Struct.\ Biol.} {\bf 9}, 594-601.

\bibitem{kim00} Kim, D.E., Fisher, C.,  \& Baker, D.  (2000) {\it  J.\ Mol.\ Biol.} {\bf 298}, 971-984. 

\bibitem{mccallister00} McCallister, E.L., Alm, E., \& Baker, D. (2000). {\it Nat.\ Struct.\ Biol.} {\bf 7}, 669-673. 

\bibitem{hamill00} Hamill, S.J., Steward, A., \& Clarke, J. (2000) {\it J.\ Mol.\ Biol.} {\bf  297}, 165-178 (2000).

\bibitem{fowler01} Fowler, S.B.,  \& Clarke, J. (2001) {\it Structure} {\bf 9}, 355-366 (2001).

\bibitem{jaeger01} J\"ager, M., Nguyen, H., Crane, J.C., Kelly, J.W., \& Gruebele, M. (2001) {\it J.\ Mol.\ Biol.} {\bf 311}, 373-393 (2001).

\bibitem{otzen02} Otzen, D.E.,  \& Oliveberg, M. (2002) {\it J.\ Mol.\ Biol.} {\bf 317}, 613-627.

\bibitem{northey02} Northey, J.G.B., Di Nardo, A.A., \& Davidson, A.R. (2002) {\it Nature Struct.\ Biol.} {\bf 9}, 126-130. 

\bibitem{garcia04} Garcia-Mira, M.M., B\"ohringer, D., \& Schmid, F.X. (2004) {\it J.\ Mol.\ Biol.} {\bf 339}, 555-569.

\bibitem{matouschek89} Matouschek, A., Kellis, J.\ T., Serrano, L., \& Fersht, A.\ R. (1989) {\it Nature} {\bf 340}, 122-126. 

\bibitem{fersht99} Fersht, A.\ R. (1999) {\it Structure and mechanism in protein science} (W. H. Freeman, New York) 

\bibitem{li94} Li, A., \& Daggett, V. (1994) {\it Proc.\ Natl.\ Acad.\ Sci.\ USA} {\bf 91}, 10430-10434 

\bibitem{li96} Li, A., \& Daggett, V. (1996) {\it J.\ Mol.\ Biol.} {\bf 257}, 412-429.

\bibitem{daggett96} Daggett, V., Li, A., Itzhaki, L.\ S., Otzen, D.\ E., \& Fersht, A.\ R. (1996) {\it J.\ Mol.\ Biol.} {\bf 257}, 430-440. 

\bibitem{ladurner98} Ladurner, A.\ G., Itzhaki, L.\ S., Daggett, V., \& Fersht, A.\ R. (1998) {\it Proc.\ Natl.\ Acad.\ Sci.\ USA} {\bf 95}, 8473-8478. 

\bibitem{lazaridis97} Lazaridis, T., \& Karplus, M. (1997)  {\it Science} {\bf 278}, 1928-1931.

\bibitem{kazmirski01} Kazmirski, S.\ L., Wong, K.-B., Freund, S.\ M.\ V., Tan, Y.-J., Fersht, A.\ R., \& Daggett, V. (2001). {\it Proc.\ Natl.\ Acad.\ Sci.\ USA} {\bf 98}, 4349-4354.

\bibitem{alm99} Alm, E., \& Baker, D. (1999) {\it Proc.\ Natl.\ Acad.\ Sci.\ USA} {\bf 96}, 11305-11310. 

\bibitem{munoz99} Mu\~noz, V., \& Eaton, W.A. (1999) {\it Proc.\ Natl.\ Acad.\ Sci.\ USA} {\bf 96}, 11311-11316. 

\bibitem{galzitskaya99} Galzitskaya, O.V.,  \& Finkelstein, A.V. (1999) {\it Proc.\ Natl.\ Acad.\ Sci.\ USA} {\bf 96}, 11299-11304. 

\bibitem{portman01} Portman, J.J., Takada, S., \& Wolynes, P.G. (2001) {\it J.\ Chem.\ Phys.} {\bf 114}, 5069-5081. 

\bibitem{alm02} Alm, E., Morozov, A.V., Kortemme, T., \& Baker, D. (2002) {\it J.\ Mol.\ Biol.} {\bf 322}, 463-476. 

\bibitem{kameda03} Kameda, T. (2003) {\it Proteins} {\bf 53}, 616-628.

\bibitem{clementi00} Clementi, C., Nymeyer, H., \& Onuchic, J.N. (2000)  {\it J.\ Mol.\ Biol.} {\bf 298}, 937-953. 

\bibitem{hoang00} Hoang, T.X., \& Cieplak, M. (2000) {\it J.\ Chem.\ Phys.} {\bf 113}, 8319-8328. 

\bibitem{li01} Li, L., \& Shakhnovich, E.I. (2001)  {\it Proc.\ Natl.\ Acad.\ Sci.\ USA} {\bf 98}, 13014-13018. 

\bibitem{karanicolas02} Karanicolas, J., \& Brooks, C.L. (2002) {\it Protein Sci.} {\bf 11}, 2351-2361. 

\bibitem{kaya03} Kaya, H., \& Chan, H.S. (2003) {\it J.\ Mol.\ Biol.} {\bf 326}, 911-931.

\bibitem{weikl04} Weikl, T.R., Palassini, M., \& Dill, K.A. (2004) {\it Protein Sci.} {\bf 13}, 822-829.

\bibitem{vendruscolo01} Vendruscolo, M., Paci, E., Dobson, C.M., \& Karplus, M. (2001) {\it Nature} {\bf 409}, 641-645.

\bibitem{lindorff03} Lindorff-Larsen, K., Paci, E., Serrano, L,  Dobson, C.M., \& Vendruscolo, M. (2003) {\it Biophys.\ J.} {\bf 85}, 1207-1214.

\bibitem{lindorff04} Lindorff-Larsen, K., Vendruscolo, M., Paci, E., \& Dobson, C.M. (2004) {\it Nature Struct.\ Mol.\ Biol.} {\bf 11}, 443-449.

\bibitem{hammond55} Hammond, G.S. (1955). {\it J.\ Am.\ Chem.\ Soc.} {\bf 77}, 334-338.

\bibitem{matthews93} Matthews, C.R. (1993).  {\it Annu.\ Rev.\ Biochem.} {\bf 62}, 653-683.

\bibitem{vanKampen} van Kampen, N.G. (1992) {\it Stochastic processes in physics and chemistry}, (Elsevier, Amsterdam) 

\bibitem{goldenberg99} Goldenberg, D.P. (1999) {\it Nature Struct.\ Biol.} {\bf 6}, 987-990.

\bibitem{munoz94a} Mu\~noz, V., \& Serrano, L. (1994) {\it J.\ Mol.\ Biol.} {\bf 245}, 275-296.

\bibitem{munoz94b} Mu\~noz, V., \& Serrano, L. (1994) {\it J.\ Mol.\ Biol.} {\bf 245}, 297-308.

\bibitem{lacroix98} Lacroix, E., Viguera, A.\ R., \& Serrano, L. (1998) {\it J.\ Mol.\ Biol.} {\bf 284}, 173-191.

\bibitem{sanchez03} S\'anchez, I.\ E.,  \& Kiefhaber, T. (2003) {\it J.\ Mol.\ Biol.} {\bf 334}, 1077-1085.

\bibitem{fersht04} Fersht, A.\ R., \& Sato, S. (2004) {\it Proc.\ Natl.\ Acad.\ Sci.\ USA} {\bf 91}, 10422-10425.

\bibitem{ozkan01} Ozkan, S.\ B., Bahar, I., \& Dill, K.\ A. (2001) {\it Nature Struct.\ Biol.} {\bf 8}, 765-769.



\end{thebibliography}
\end{document}